\documentclass[12pt,preprint]{aastex}
\shortauthors{}
\shorttitle{}

\begin{document}

\title{Molecular cloud chemistry and the importance of dielectronic recombination}

\author{P. Bryans,\altaffilmark{1}
	H. Kreckel,\altaffilmark{1}
	E. Roueff,\altaffilmark{2}
	V. Wakelam\altaffilmark{3}
	and D. W. Savin,\altaffilmark{1}}
\altaffiltext{1}{Columbia Astrophysics Laboratory, Columbia University,
550 West 120th St, MC 5247, New York, NY 10027-6601}
\altaffiltext{2}{LUTH and UMR 8102 du CNRS, Observatoire de Paris, 
Section de Meudon, Paris, France}
\altaffiltext{3}{Universit\'{e} Bordeaux, Laboratoire d'Astrophysique
de Bordeaux, CNRS/INSU, UMR 5804, BP89 33271 Floirac Cedex, France}

\begin{abstract}

Dielectronic recombination (DR) of singly charged ions is a reaction pathway
that is commonly neglected in chemical models of molecular clouds. In this study
we include state-of-the-art DR data for He$^+$, C$^+$, N$^+$, O$^+$, Na$^+$, and
Mg$^+$ in chemical models used to simulate dense molecular clouds, protostars,
and diffuse molecular clouds.  We also update the radiative recombination (RR)
rate coefficients for H$^+$, He$^+$, C$^+$, N$^+$, O$^+$,   Na$^+$, and Mg$^+$
to the current state-of-the-art values.   The new RR data has little effect on
the models. However, the inclusion of DR results in significant differences  in
gas-grain models of dense, cold molecular clouds for the evolution of a number
of surface and gas-phase species.   We find differences of a factor of 2 in the
abundance for 74 of the 655 species at times of $10^4$--$10^6$~years
in this model when we include DR. Of these
74 species, 16 have at least a factor of 10 difference in abundance.  We find
the largest differences for species formed on the surface of dust grains.  These
differences are due primarily to the addition of C$^+$ DR, which increases the
neutral C abundance, thereby enhancing the accretion of C onto dust.  These
results may be important for the warm-up phase of molecular clouds when surface
species are desorbed into the gas phase.  We also note that no reliable
state-of-the-art RR or DR data exist for Si$^+$, P$^+$, S$^+$, Cl$^+$, and
Fe$^+$. Modern calculations for these ions are needed to better constrain
molecular cloud models.

\end{abstract}

\keywords{astrochemistry --- atomic data --- atomic processes ---
ISM: atoms --- dust, extinction --- ISM: molecules} 

\section{Introduction}
\label{sec:intro}

Investigating the physics and chemistry of molecular clouds (also known as
quiescent cores) is crucial if one is to understand the processes that
ultimately lead to star formation. Such studies are also important in the field
of astrobiology since these clouds are the birthplace of the first organic
molecules. 

Chemical models used to describe the evolution of molecular clouds
typically include hundreds of species.   At molecular cloud temperatures
($\lesssim 100$~K) these atoms and molecules are either neutral or singly
ionized. The abundances of these species are, in turn, governed by thousands of
reactions that are highly non-linear. These derived abundances are sensitive to
the accuracy of the rate coefficients involved. The implication of uncertainties
in the rate coefficients used in chemical models has been investigated by
\citet{Roue96a}, \citet{Vasy04a,Vasy08a}, \citet{Wake05a,Wake06a},
and others. However, these
studies cannot account for reaction processes that are not included in the
models in the first place.

To the best of our knowledge, the only  recombination process of free electrons
with atomic ions included in any molecular cloud simulation is radiative
recombination (RR). Recent calculations by 
\citet{Badn06a}\footnote{http://amdpp.phys.strath.ac.uk/tamoc/RR/} have improved
the accuracy of the rate coefficients of these reactions for a number of ions.
The alternative pathway of dielectronic recombination (DR) has previously not
been included in molecular cloud models. This is largely due to the bulk of
published DR calculations and experiments being valid only for plasmas of higher
temperature. Recently, however, theoretical
calculations by \citet{Badn03a}\footnote{http://amdpp.phys.strath.ac.uk/tamoc/DR/} 
have probed DR at the low temperature
regimes of molecular clouds. At these temperatures, the DR rate coefficient can
be a factor of 5 or more greater than the RR rate coefficient for certain
systems.

In the present paper, we both update the RR data and include DR for the relevant
singly charged atomic ions.  We do this for several molecular cloud models under
a variety of initial conditions.  We use  Nahoon \citep{Wake04a} to simulate
dense clouds and protostars, the OSU gas-grain code \citep{osu,Garr06a} to simulate
dense clouds,  and the Meudon Photodissociation Region (PDR) code 
\citep{pdr,lepe06a} to
simulate both diffuse and dense PDRs. 
We compare the species abundances with those
calculated with and without DR included  in order to show the effects on the
models. The remainder of the paper is organized as follows: In
Sec.~\ref{sec:recom} we review the recent developments in our understanding RR
and DR. Section~\ref{sec:models} outlines the chemical models we use to simulate
dense clouds, diffuse clouds, and protostars. In Sec.~\ref{sec:results} we
present the results of including DR in these chemical models. 
Concluding remarks are given in
Sec.~\ref{sec:conclusion}.

\section{Recombination data}
\label{sec:recom}

\subsection{Radiative Recombination}
\label{sec:rr}

Radiative recombination (RR) is a one-step recombination process that occurs
when a free electron is captured by an ion. Energy and momentum are conserved in
the process by the simultaneous emission of a photon. At the low temperatures
typical of molecular clouds, RR is the dominant ion-electron recombination
process for most ions. Until now, the molecular cloud codes that we use in the
present work (see Sec.~\ref{sec:models}) have used RR rate coefficients from the
UMIST  database\footnote{http://www.udfa.net/} \citep{Wood07a}. For the most
part it is unclear where the RR data in the UMIST database stem from as there is
often no reference given; this is the case for the rate coefficients of  H$^+$,
He$^+$, Na$^+$, Mg$^+$, Si$^+$, P$^+$, S$^+$, Cl$^+$, and Fe$^+$.  There is no
RR data for F$^+$. For C$^+$ and N$^+$ the rate coefficients are those of
\citet{Naha97a} and for O$^+$ the rate coefficient is from \citet{Naha99a}. The
data of Nahar \& Pradhan are actually unified RR+DR calculations using
LS-coupling. As a result, they have no DR component at the low temperatures of
molecular clouds since they do not account for fine structure transitions of the
ground term (see Sec.~\ref{sec:dr}).

In recent years there have been attempts to better understand the RR process
using state-of-the-art computational techniques. \citet{Badn06a}
 has calculated
RR rate coefficients for all elements from H through Zn for bare through Mg-like
isoelectronic sequences. We have
implemented these calculations for
the ions included in the chemical networks of the models considered here, namely
H$^+$, He$^+$, C$^+$, N$^+$, O$^+$, Na$^+$, and Mg$^+$. For the ions that \citet{Badn06a} has not
calculated---namely Si$^+$, S$^+$, Cl$^+$, and Fe$^+$---we have used the
rate coefficients recommended by \citet{Mazz98a}. These come from
\citet{Aldr73a} for Si$^+$ and S$^+$; an extrapolation of the S-like sequence is
used to estimate the  Cl$^+$ rate coefficient; and the rate coefficient from
Schull \& van Steenberg (1982; refitted with the Verner \& Ferland 1996 formula)
is used for Fe$^+$.

\begin{figure}
  \centering
  \includegraphics[width=15cm]{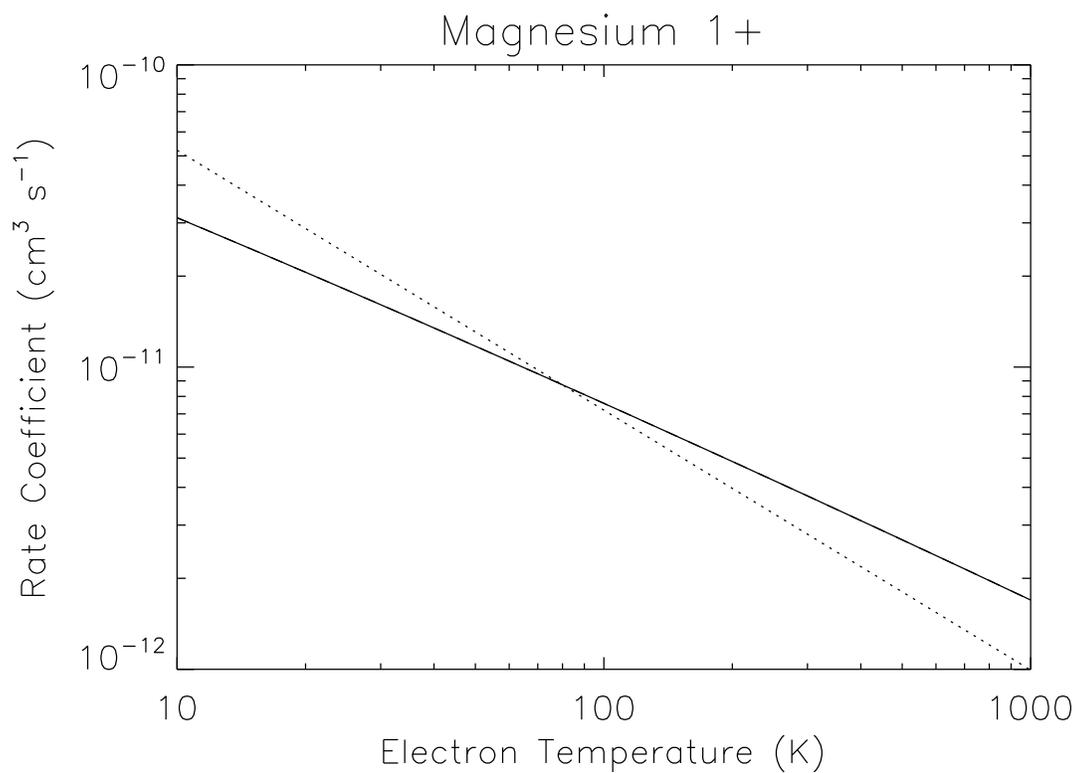}
  \caption[]{The RR rate coefficient for Mg$^+$ forming Mg.
  The dotted line is the rate coefficient from the UMIST database and
  the solid line shows the modern calculation of \citet{Badn06a}.}
  \label{fig:rr}
\end{figure}

At molecular cloud temperatures, the largest difference found between the RR
rate coefficients of the UMIST dataset and the \citet{Badn06a} RR data is for
Mg$^+$, as is shown in Fig.~\ref{fig:rr}. At a temperature of 10~K, typical of a
cold molecular cloud, the UMIST rate coefficient is 60\% larger than that of
Badnell.

\subsection{Dielectronic recombination}
\label{sec:dr}

Dielectronic recombination (DR) is a two-step recombination process that begins
when a free electron collisionally excites a core electron of an ion and is
simultaneously captured. The core electron excitation can be labeled $nl_j\to
n^\prime l^\prime_{j^\prime}$, where $n$ is the principal quantum number, $l$
the orbital angular momentum, and $j$ the total angular momentum.  The energy of
this intermediate system lies in the continuum and the complex may autoionize.
The DR process is complete when the system emits a photon, reducing the total
energy of the recombined system to below its ionization threshold.  Conservation
of energy requires that for DR to go forward   
\begin{equation} 
E_k=\Delta E-E_b. 
\end{equation} 
Here $E_k$ is the kinetic energy of the incident electron,
$\Delta E$ the excitation energy of the initially bound electron in the 
presence of the captured electron, and $E_b$ the binding energy released when
the incident electron is captured onto the excited ion. Because $\Delta E$ and
$E_b$ are quantized, DR is a resonant process occurring for a given channel at
energies $E_k<\Delta E$.

For a molecular cloud of temperature $T_{\rm cloud}$, the important DR channels
are those where $\Delta E\sim k_BT_{\rm cloud}\lesssim 0.01$~eV.
Thus, from atomic energy structure considerations alone, it is clear that
fine structure excitations of the ground term
are important in molecular clouds.
Until recently, the
preponderance of DR data have been calculated using LS-coupling.
However, this coupling scheme does not include these low-energy
resonances.
Recent state-of-the-art calculations by 
\citet{Badn03a}, using
intermediate coupling,
have accounted for these  fine structure channels, calculating
DR rate coefficients for H-
through Mg-like ions of all elements from He through Zn.

\begin{figure}
  \centering
  \includegraphics[width=15cm]{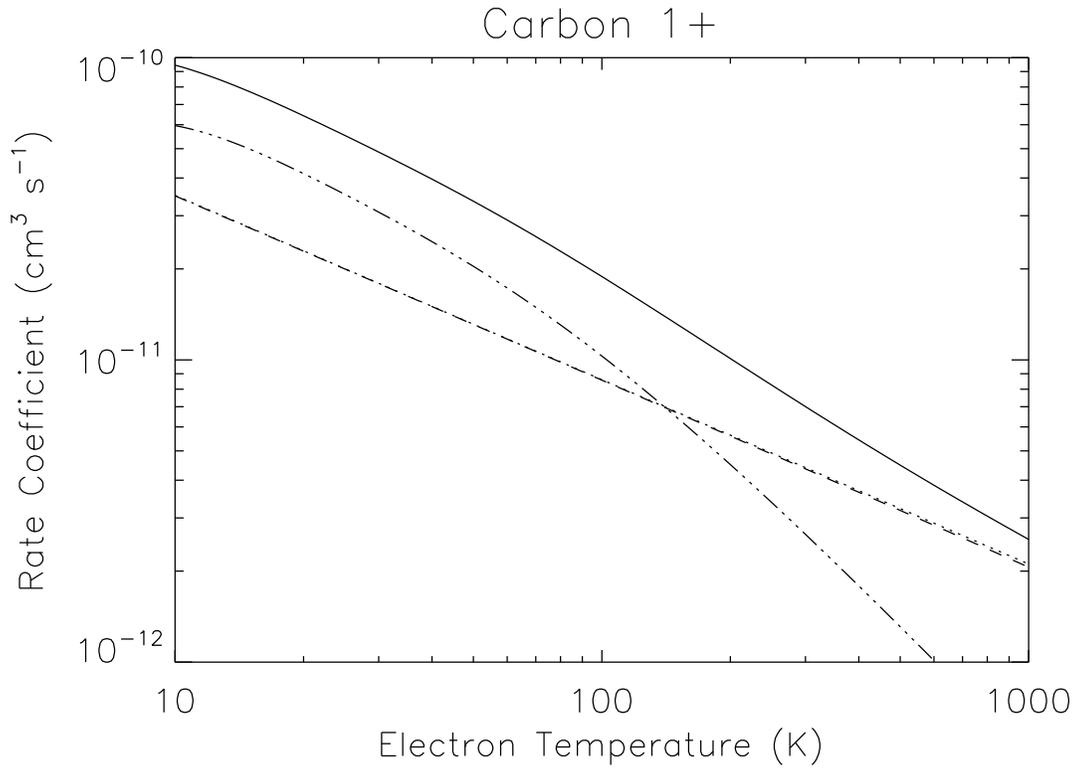}
  \caption[]{The DR and RR rate coefficients for C$^+$ forming C.
  The dotted line is the RR rate coefficient from the UMIST database.
  The dashed line shows the modern  RR calculation 
  of \citet{Badn06a} and lies almost exactly on top of the
  UMIST data.  The dot-dot-dot-dashed line shows the DR calculation
  of \citet{Badn03a}.
  The solid line is the total RR+DR recombination rate coefficient
  of Badnell.}
  \label{fig:dr c}
\end{figure}

\begin{figure}
  \centering
  \includegraphics[width=15cm]{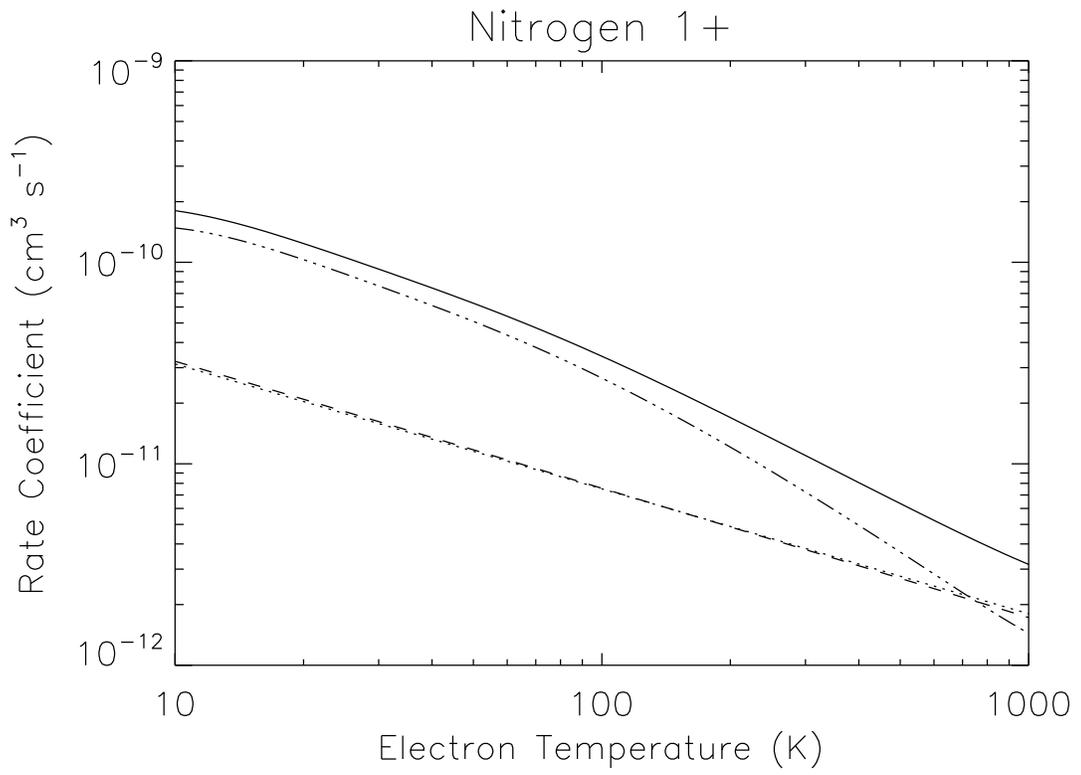}
  \caption[]{Same as Fig.~\ref{fig:dr c} but for N$^+$ forming N.}
  \label{fig:dr n}
\end{figure}

For the singly-ionized ions of interest in molecular  clouds---H$^+$, He$^+$,
C$^+$, N$^+$, O$^+$, F$^+$, Na$^+$, Mg$^+$, Si$^+$, P$^+$, S$^+$, Cl$^+$, and Fe$^+$---
one would expect a significant low-temperature DR contribution for C$^+$, N$^+$,
F$^+$, Si$^+$, P$^+$, Cl$^+$ and Fe$^+$ due to their fine structure splitting of the
ground term. The other ions either do not undergo DR (i.e., H$^+$) or have no
such fine structure splitting. Of these five ions with fine structure
splitting,  \citet{Badn03a} has calculated DR only for C$^+$, N$^+$, and F$^+$.
Figs.~\ref{fig:dr c} and \ref{fig:dr n} show
 a comparison of the DR and RR rate coefficients for
C$^+$ and N$^+$, respectively. 
At temperatures $\lesssim 100$~K, typical of molecular clouds, the DR
component dominates the electron-ion recombination rate coefficient highlighting the
importance of including this recombination process. For the present work we have
implemented the DR rate coefficients of \citet{Badn03a} for He$^+$, C$^+$,
N$^+$, O$^+$, F$^+$, Na$^+$, and Mg$^+$.
No such non-LS coupling calculations exist for Si$^+$, P$^+$, Cl$^+$ and Fe$^+$.

\section{Models}
\label{sec:models}

\subsection{Nahoon}
\label{sec:nahoon}

The Nahoon code \citep{Wake04a} is a pseudo-time-dependent chemical model that
computes the chemical evolution of gas-phase species for a fixed gas temperature
and density.   It is suited for modeling dense molecular clouds and protostars.
The model includes 452 species with initial conditions being atomic (neutral and
singly charged) He, C, N, O, F, Na, Mg, Si, P, S, Cl, and Fe plus molecular
hydrogen.   Dust grains are included in the model but are not fully treated;
ions are allowed to neutralize by charge exchange with negative grains but
accretion of species onto grains is not included, and thus no grain surface
chemistry is possible. The code computes the evolution of the abundance of each
species as governed by the 4423 included reactions  between species. The reaction rate
coefficients come  from the 
osu.2005\footnote{http://www.physics.ohio-state.edu/$\sim$eric/research\_files/osu.2005}
chemical network \citep{Smit04a}. The RR rate coefficients in this network are
those from  the UMIST database (see discussion in Sec.~\ref{sec:rr}). We have
replaced these RR rate coefficients with those of \citet{Badn06a} for H$^+$,
He$^+$, C$^+$, N$^+$, O$^+$,  Na$^+$, and Mg$^+$. We have also added the DR
reactions for He$^+$, C$^+$, N$^+$, O$^+$, Na$^+$, and Mg$^+$ using the rate
coefficients of \citet{Badn03a}. For the other ions that have not been
calculated by  \citet{Badn03a} or \citet{Badn06a}---Si$^+$, S$^+$, Cl$^+$, and Fe$^+$---we
have used the RR and DR rate coefficients of \citet{Mazz98a}. While the elements
F and P are present in Nahoon, recombination of F$^+$ and P$^+$ are not
included.  We have chosen not to introduce these processes and instead
concentrate on the effect of the updated RR and DR rate coefficients for those
reactions that are already present.

With these new reactions included, we have used Nahoon to simulate
the evolution of dense cold clouds and protostars.
A dense cold cloud is simulated by assuming a temperature of 10~K, an
initial H$_2$ density of $10^4$~cm$^{-3}$, a visual extinction of 10
so that the photochemistry driven by external UV photons does not occur, and
a fixed cosmic-ray ionization rate of $1.3\times 10^{-17}$~s$^{-1}$.
These dense cloud conditions are given in Table~\ref{tab:initial}.
We use the low-metal elemental abundances of \citet{Grae82a} as
our initial conditions.  These are also given in Table~\ref{tab:initial}.
For a protostar, we take our initial gas-phase chemical composition
as that 
computed by Nahoon for the above dense cloud conditions at $10^5$~years.
Additionally, we
increased the temperature to 100~K, the
initial H$_2$ density to $10^7$~cm$^{-3}$, and
set the abundances of H$_2$O, H$_2$CO, CH$_4$O, and CH$_4$
as given in Table~\ref{tab:initial}.
This is required because Nahoon does not include the surface chemistry
that is needed to form these species, so we artificially increase their
abundance in the gas phase to simulate protostar conditions.
Results for both dense clouds and protostars
 are presented in Sec.~\ref{sec:results}.

\subsection{OSU gas-grain code}
\label{sec:osu}

The OSU gas-grain code \citep{osu,Garr06a} is similar to Nahoon in that it follows the
evolution of the species abundances as a function of time for a fixed cloud
temperature and density.  Like Nahoon, it is used to simulate the conditions of
dense molecular clouds. However, unlike Nahoon, the OSU gas-grain code includes
surface chemistry in addition to gas-phase chemistry. This allows for
time-dependent accretion onto grain surfaces as well as
thermal and cosmic ray-induced evaporation
from the surfaces. 
The code assumes a sticking coefficient of unity for neutral species that
strike
grains.  It is unclear whether charged species can also accrete onto grains
\citep{Wats76a}.  It is assumed here that they do not, with the
energy released on neutralization resulting in the species
desorbing from the grain surface.
This model also includes the new non-thermal evaporation mechanism of
\citet{Garr07a}.  This process assumes that the energy released by
exothermic surface reactions partially evaporate the products.

Species can react on the grain surfaces to form
 molecules  more efficiently than is possible solely in the gas phase.
This surface chemistry is an important contributor to the species
present in the warm-up phase of hot molecular cores \citep{Garr06a}.
Species that form on the grains at earlier times are injected into
the gas phase when the dust is heated by a nearby protostar.

The OSU code uses the osu.2005 chemical network, although there is no F
present in the OSU gas-grain code, unlike in Nahoon. 
With the addition of the surface species to the network used with
Nahoon, there are 655 species and 6309 reactions.
We added
the new RR and DR rate coefficients in the same way as was done in Nahoon
(see Sec.~\ref{sec:nahoon}).
With these changes to the chemical network, we have run the OSU code
with the same initial conditions as used in Sec.~\ref{sec:nahoon}
when simulating a dense cold cloud (see Table~\ref{tab:initial}).
We present our results in Sec.~\ref{sec:results}.

\subsection{PDR code}
\label{sec:pdr}

In order to simulate a molecular cloud with a significant impingent 
radiation field, we use the PDR code of the Meudon  group
\citep{pdr,lepe06a}\footnote{http://aristote.obspm.fr/MIS/pdr/pdr1.html}.
This code is used to simulate clouds with no rapidly evolving processes
occurring.   The
cloud is treated in steady state as an infinite slab of gas and dust irradiated
by an ultraviolet radiation field impinging on both sides of the cloud.
While dust is included in the model, there is no 
surface chemistry other than allowing for the formation of H$_2$
which cannot be produced rapidly enough in the gas phase.

Using a standard chemistry file with 120 species
we have added the new RR and DR data to the PDR code.
The elemental abundances are given in Table~\ref{tab:initial}.
We have run the model with physical conditions that simulate
a diffuse PDR and, separately, a dense PDR.
For diffuse conditions, we set the 
H$_2$ density to $25$~cm$^{-3}$, the temperature to
20~K, and the cosmic ray ionization rate of $5\times 10^{-17}$~s$^{-1}$.
The radiation field is 1~Draine at the edges of the cloud.
For a dense PDR, we simulate the Horsehead nebula \citep{Pety05a}
with an H$_2$ density of $5\times10^4$~cm$^{-3}$, a temperature of 90~K, and
a cosmic ray ionization rate of $5\times 10^{-17}$~s$^{-1}$.
For this case the radiation field is 100~Draine at the edge of the cloud.
The species abundances are calculated as a function of 
depth into the cloud.
Results are presented in Sec.~\ref{sec:results}.

\section{Results}
\label{sec:results}

For each of the models described in Sec.~\ref{sec:models} we have run the 
simulation with the original chemical network without any additions, with just
the new RR, and again with the new RR and DR rate coefficients included. When
running the models with only the new RR data included we  find no differences
greater than a factor of 2 for any of the models. Only for the OSU gas-grain
code do we find abundance differences of greater than 50\%.  These
are found for gas-phase Fe, surface
FeH, and surface MgH$_2$, resulting from
the change in the Mg$^+$ and Fe$^+$ RR rate coefficients. The more significant
results reported in the remainder of this section can thus be attributed primarily
to the effect of DR on the models.

\begin{figure}
  \centering
  \includegraphics[width=15cm]{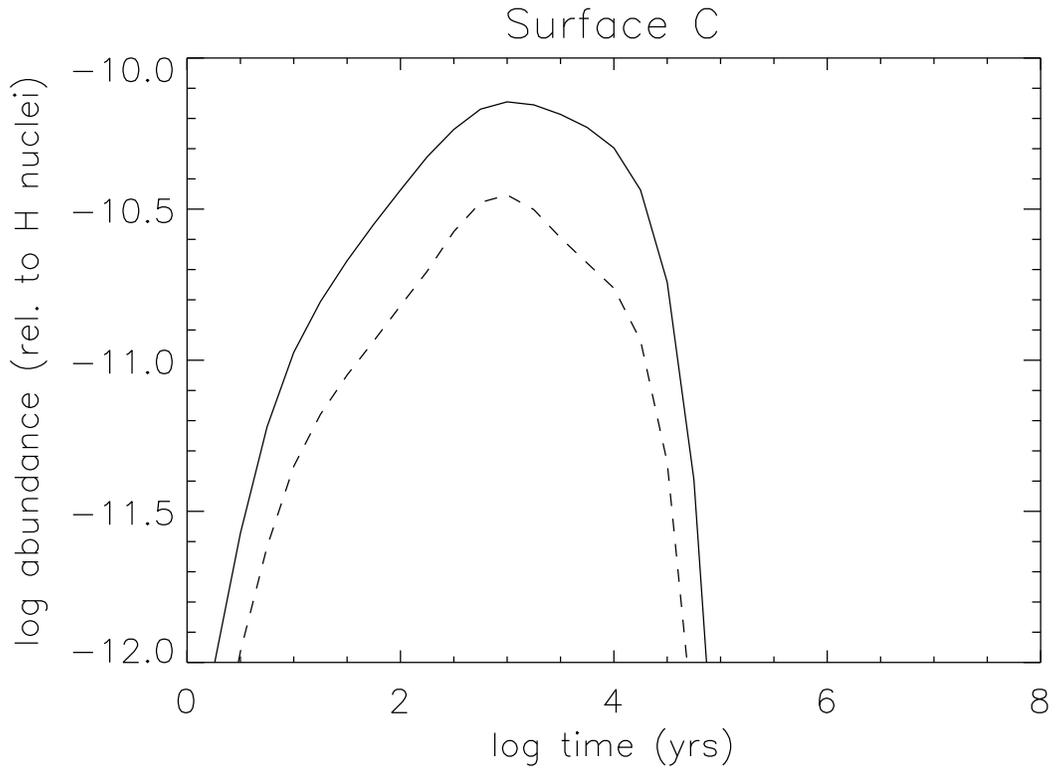}
  \caption[]{The abundance of surface C with respect to the total H
  nuclei density as a function of time as
  calculated by the OSU gas-grain code.  The solid line is with
  the new RR and DR data included in the model
  and the dotted line is without the new RR and DR included.}
  \label{fig:jc}
\end{figure}

\begin{figure}
  \centering
  \includegraphics[width=15cm]{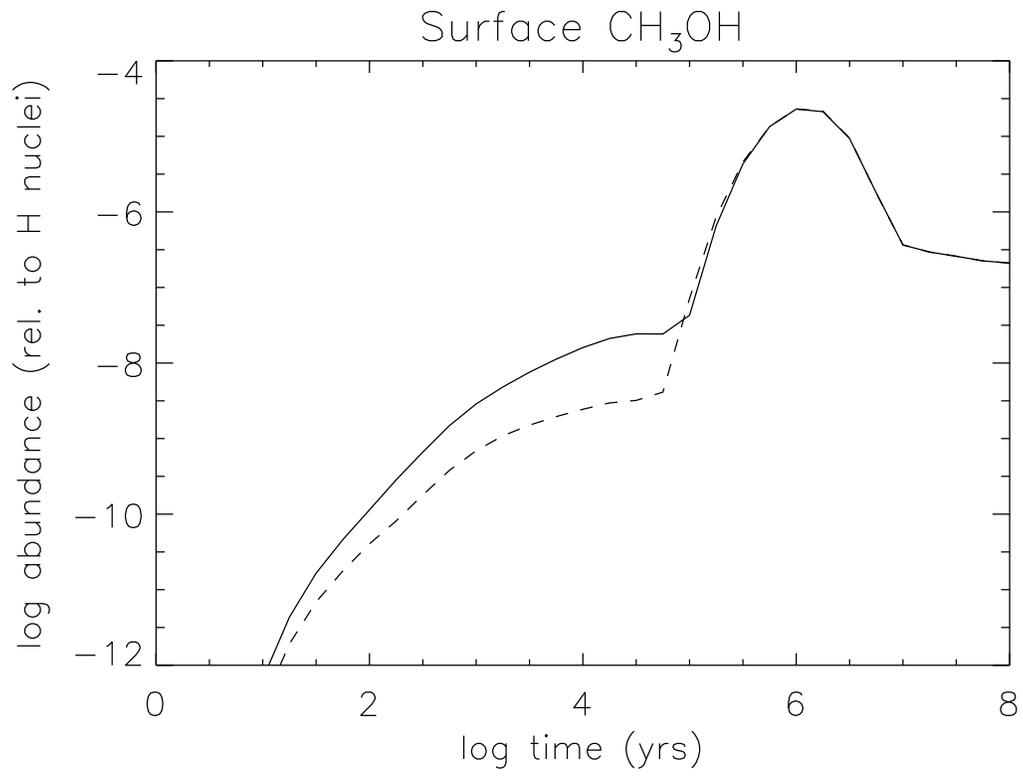}
  \caption[]{Same as Fig.~\ref{fig:jc} but for surface CH$_3$OH.}
  \label{fig:jch3oh}
\end{figure}

\begin{figure}
  \centering
  \includegraphics[width=15cm]{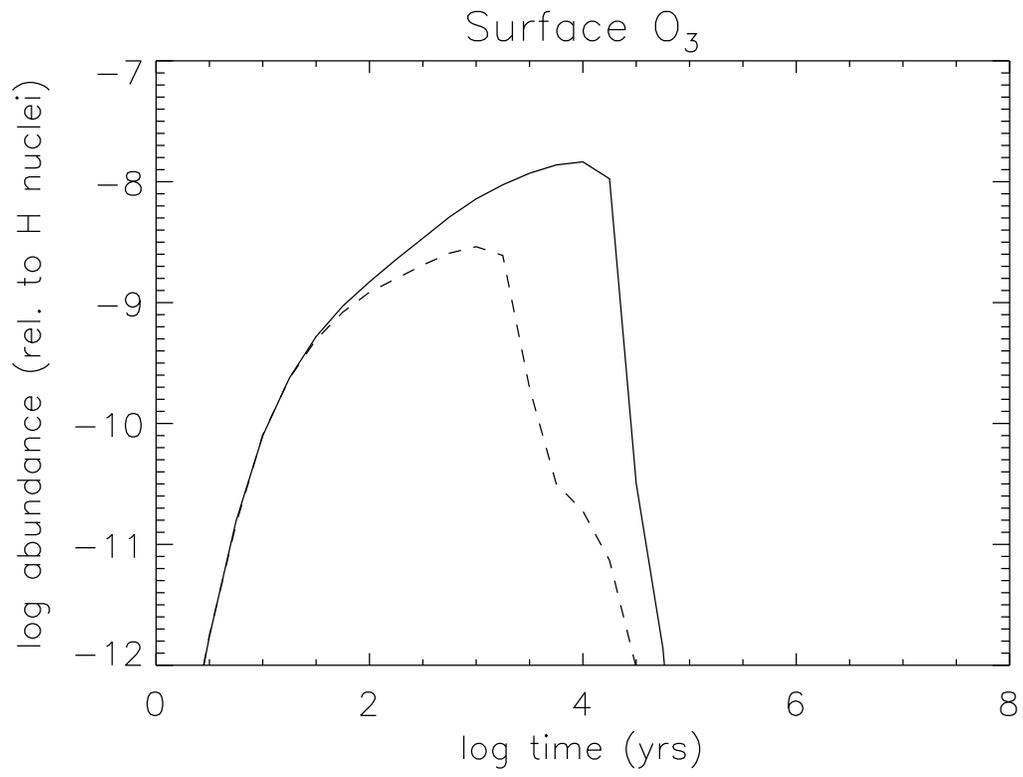}
  \caption[]{Same as Fig.~\ref{fig:jc} but for surface O$_3$.}
  \label{fig:jo3}
\end{figure}

Simulations of a dense, cold molecular cloud of low metallicity were run with
the OSU gas-grain code, which computes both gas-phase and surface chemistry. We
find that the introduction of DR results in significant differences in certain
species evolution.   To identify those differences that are most important in
the cloud chemistry we consider differences only at times when species have an abundance  (either
before or after the inclusion of DR) of at least $10^{-12}$ with respect to the
total H nuclei density.
This represents those species that are most likely to  be detectable.
Secondly, we only consider abundances at times of $10^4$--$10^6$~years in the
model evolution as it is during this period in the evolution of the model that
observed molecular cloud conditions are best represented. With these criteria,
we find 100 species to have a difference in abundance of at least  50\% on
including DR.  We list in Table~\ref{tab:abundances} the 74 species for which we
find at least a factor of 2 difference in abundance. Of these 74 species 48 are
attached to the surface of dust grains and 26 are found in the gas phase.   The
effect of the new DR rate coefficients on  species abundances is over 2 orders
of magnitude for 16 species. The greatest effect is found for surface O$_3$,
with a 3 orders of magnitude difference in abundance found. In
Figs.~\ref{fig:jc}, \ref{fig:jch3oh}, and \ref{fig:jo3} we show the evolution of
the abundance of surface C, surface CH$_3$OH, and surface O$_3$, respectively.

We have run the gas-grain model a number of times, including  DR  for only a
single selected ion in each run, and repeated this for every ion. This allows us
to determine which ion has the largest effect on the differences we have found. 
These results indicate that the abundance differences listed in
Table~\ref{tab:abundances} are primarily due to the introduction of the C$^+$
DR.  When the C$^+$ DR rate coefficient is not included in the model we find no
differences in abundance greater than a factor of 2. The inclusion of DR
increases the C$^+$ recombination rate, neutralizing C earlier in the model
evolution.  The model allows neutral species to accrete onto grains but not
charged species.  Thus, with an increased abundance of neutral C, the surface
chemistry is enriched and results in the abundance differences that we observe.

Despite the large increase in the electron-ion recombination rate
coefficient for N$^+$ (Fig.~\ref{fig:dr n}),
the comparatively small effect of the inclusion of N$^+$ DR can be
explained by considering the ionization balance of N in the simulation.
Initially (see Table~\ref{tab:initial}) all N in the model is in
the neutral charge state, so recombination of N$^+$ with a free electron
is not possible.  As the model evolves, some N$^+$ is formed but
neutral N remains many orders of magnitude greater in abundance.

\begin{figure}
  \centering
  \includegraphics[width=15cm]{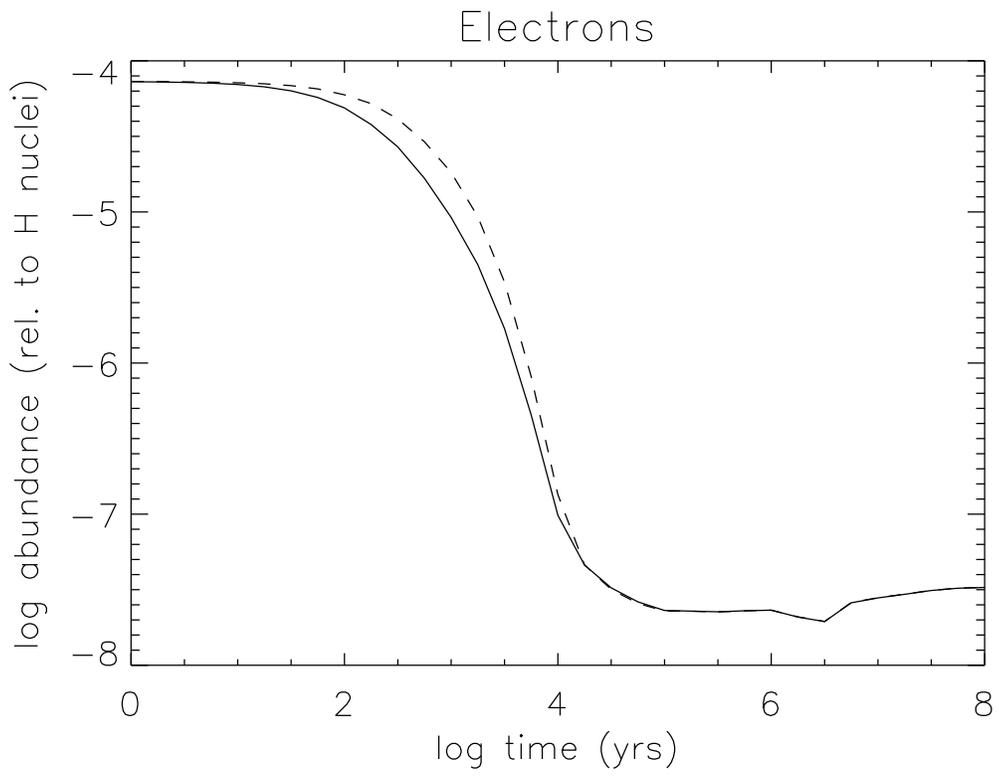}
  \caption[]{Same as Fig.~\ref{fig:jc} but for free electrons.}
  \label{fig:elec}
\end{figure}

In contrast, atomic C, due to its lower first ionization potential, is initially
all singly ionized by UV radiation shortward of the 13.6~eV ionization
potential of H but above the 11.3~eV ionization potential of C (again see
Table~\ref{tab:initial}). Thus, recombination of C$^+$ is important from the
very start of the model evolution and the inclusion of DR greatly increases the
electron-ion recombination rate. The increase in the C recombination rate also results in a
change in the ionization fraction of the gas as can be seen in the abundance
evolution of the sum of all negatively charged species. To show this we plot the
free electron abundance in Fig.~\ref{fig:elec}.  The abundance of all other
negatively charged species are negligible in comparison.

To investigate the importance of the surface chemistry on our results, we have
run the gas-grain model with no accretion of species onto dust grains allowed. 
We have compared results with and without DR included. With the surface
chemistry excluded, we find differences in abundance of 50\% for 32 species but
factor of 2 increases for only 5 species---CH$_3$CN, CH$_3$CO$^+$,
HC$_2$NC, C$_3$H$_3^+$, and HC$_3$N. These 5 gas-phase molecules had among the
smallest changes of the 74 species we identified as having a factor
of 2 difference when the surface chemistry was included. Of the 74 species listed
in Table~\ref{tab:abundances}, 48 are surface species which are unable to form
when we do not include the surface chemistry. However, there are also 26
gas-phase species in Table~\ref{tab:abundances}. Only the 5  gas-phase species
identified above have differences of a factor of 2 when the surface chemistry is
excluded. Of the 21 other gas-phase species in Table~\ref{tab:abundances}, 11
have no formation routes without the surface chemistry included.  The other 10,
whose abundances are greatly enhanced by formation routes on grain surfaces,
have changes of less than 50\% in a purely gas-phase model.
Thus, the majority, and indeed the largest, abundance differences seen when
including DR in the OSU gas-grain model are found for species influenced by
surface reactions.

Not surprisingly then,  we do not find significant abundance differences 
when running the dense cloud simulation using Nahoon,
which does not include grain surface chemistry.  All species in Nahoon are in
the gas phase. The introduction of the DR data resulted in 
20 species having an abundance difference of 50\% and
only 2 species, OCS
and C$_2$H$_3$N, having abundance increases of greater than a factor of 2.
When running the protostar
simulation using Nahoon we again find few significant differences when DR is
included; 21 species have an abundance difference of 50\% or more. 
The only species with greater than a factor of 2 difference are
C$_2$S,     HCN, HNC, C$_2$H$_3$N, HCOOCH$_3$, CH$_3$OCH$_3 $, H$_2$CN$^+$, and 
CH$_3$OCH$_4^+$, all of which are increased in abundance with DR relative
to without.

\begin{figure}
  \centering
  \includegraphics[width=15cm]{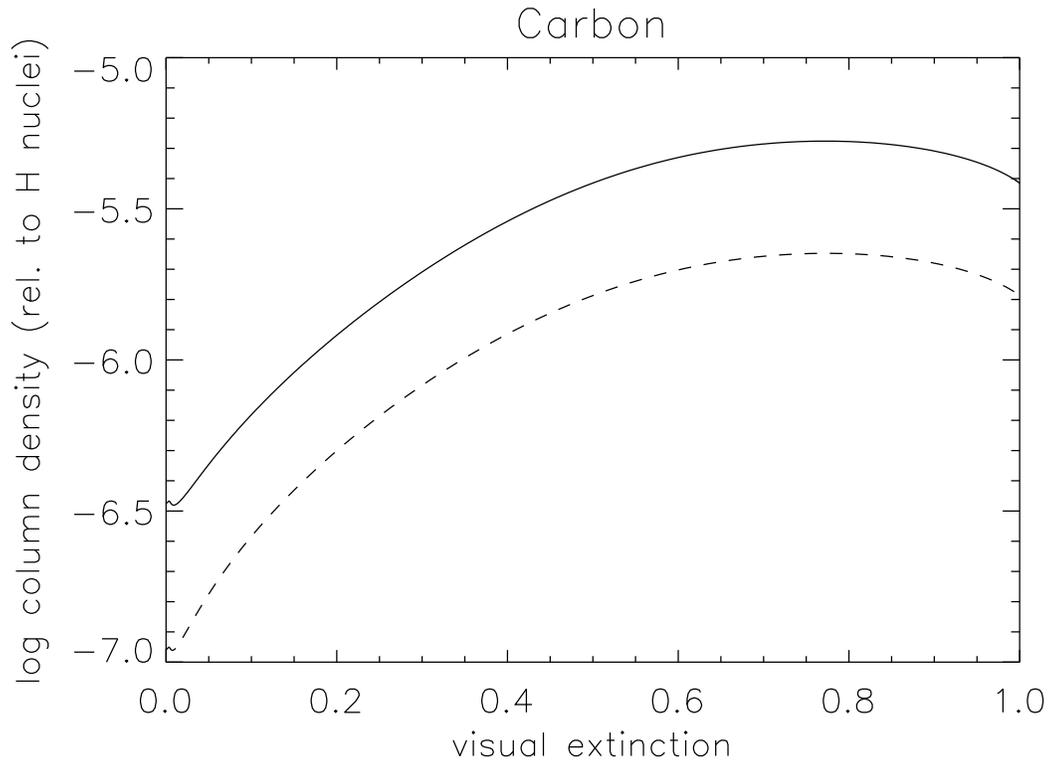}
  \caption[]{The column density of C relative to the total H nuclei column density 
  as a function of visual extinction
  (i.e., depth into the cloud) calculated by the Meudon PDR code
  for a diffuse PDR.  
  The solid line is with the new RR and
  DR included in the model and the dashed line is without
  the new RR and DR included.}
  \label{fig:c pdr diffuse}
\end{figure}

\begin{figure}
  \centering
  \includegraphics[width=15cm]{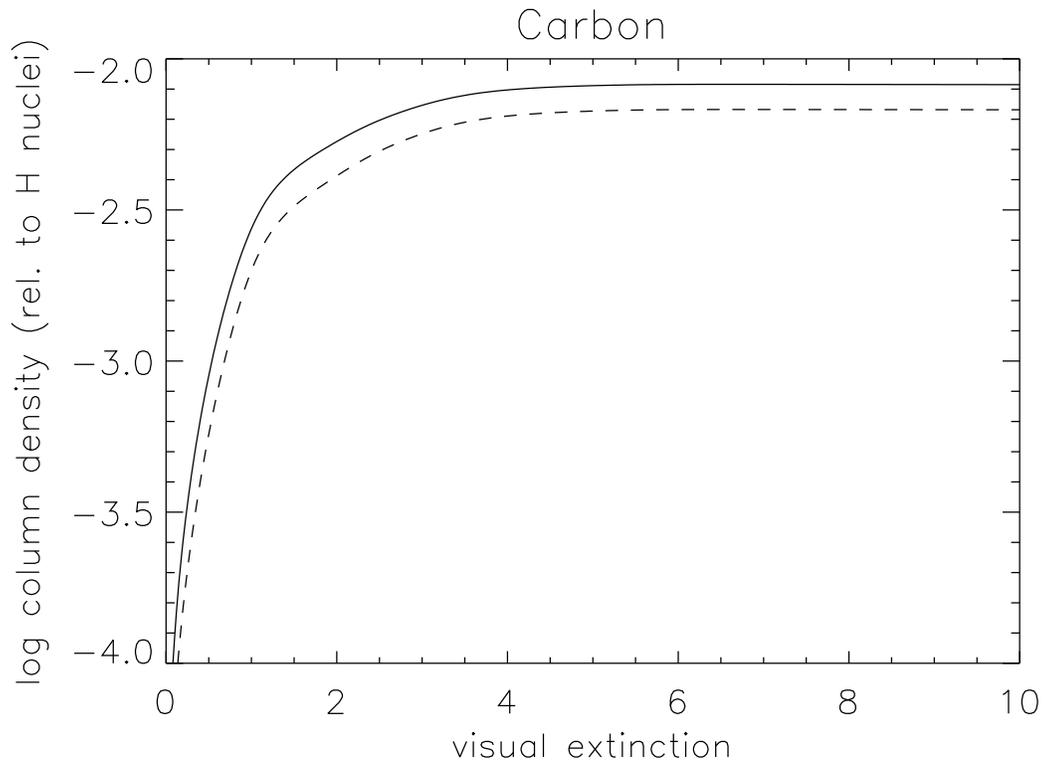}
  \caption[]{Same as Fig.~\ref{fig:c pdr diffuse} but for a dense PDR.}
  \label{fig:c pdr dense}
\end{figure}

Simulations of diffuse and dense PDRs
were carried out using the Meudon PDR code,
allowing us to probe the effect of DR in a molecular cloud with a
radiation field present.
On introducing the new  DR rate coefficients there is a large
increase in the total recombination rate of C$^+$ and we see
a significant effect on the abundance of neutral carbon as a result.
This is the case for both diffuse and dense PDRs.
The column density of C is shown in Figs.~\ref{fig:c pdr diffuse}
and \ref{fig:c pdr dense} for diffuse and dense regions, respectively,
as a function of visual extinction (i.e., depth into the cloud).
However, we find no significant abundance differences for any other
species in the diffuse case.
For a dense PDR, 
there are only 5 other species that show an abundance difference
greater than 50\%---SO$_2$, O$_2$H$^+$, HSO$_2^+$, HOCS$^+$, and C$_4$.
These species all increase in abundance with the introduction of DR
but the increases are all less than a factor of 2.

\section{Conclusions}
\label{sec:conclusion}

This work has investigated the importance of new RR and DR rate coefficients
in molecular cloud models.
We have updated the RR
rate coefficients of H$^+$, He$^+$, C$^+$, N$^+$, O$^+$, 
Na$^+$, and Mg$^+$ to the current state-of-the-art
in 3 different molecular cloud chemical models.
We have run these models under different conditions and
compared the abundances of the species present with and without
the new RR rate coefficients.
We find that the new RR data have no significant effects on
the model results (i.e., less than a factor of 2 effect).
We have also included DR for He$^+$, C$^+$, N$^+$, O$^+$, 
Na$^+$, and Mg$^+$ in the 3 chemical models, finding some
sizable differences (greater than a factor of 2)
in the evolution of certain species.

We find that DR has the greatest effect  for  dense cloud models, particularly
when surface chemistry is included. Of the models we ran here, only the OSU
gas-grain code includes the extensive surface chemistry that allows species to be accreted
onto grains and subsequently react with other surface species.  Nahoon
does not include this surface chemistry and the Meudon code only allows the
formation of H$_2$ on surfaces. DR has its greatest
influence for species that form on grain surfaces.   These species may be
particularly important in the warm-up phase of molecular clouds when they become
desorbed into the gas phase. We find DR of C$^+$  to be primarily responsible
for the abundance changes in the model.  This is because charged species are
assumed not to accrete onto grains and adding DR increases the gas-phase neutral
C abundance available for first accreting onto grains and then participating in
the surface chemistry.

We conclude by noting that there are some ions for which no state-of-the-art RR
or DR data exist, specifically Si$^+$, P$^+$, S$^+$, Cl$^+$, and Fe$^+$. Given
the fine structure present in the ground terms of Si$^+$, Cl$^+$, P$^+$, and
Fe$^+$ one would expect a significant DR component to the total electron-ion recombination
rate coefficient at low temperatures for these ions.  These 4 elements also have
first ionization potentials below that of H (13.6~eV). Given that they would
thus be ionized in molecular clouds, the effect of increasing their
recombination rates may be important.  Modern calculations of both RR and DR
rate coefficients for these ions are needed for generating reliable molecular
cloud chemical models.

\acknowledgments

\begin{deluxetable}{ccccc}
\tablewidth{0pt}
\tablecaption{Initial species abundances relative to the total H
nuclei density, and physical conditions
for dense clouds, protostars, and dense and diffuse PDRs.}
\tablehead{\colhead{} & \multicolumn{4}{c}{Initial conditions} \\
\cline{2-5} \colhead{Parameter} & \colhead{Dense Cloud} & 
\colhead{Protostar\tablenotemark{1}} &
 \colhead{Diffuse PDR} & \colhead{Dense PDR}}
\startdata
\label{tab:initial}
H       & --                  & --                   & $9.90\times10^{-1}$ & $9.90\times10^{-1}$   \\
He      & $1.40\times10^{-1}$ & $1.40\times10^{-1}$  & $1.00\times10^{-1}$ & $1.00\times10^{-1}$   \\
N       & $2.14\times10^{-5}$ & $2.14\times10^{-5}$  & $7.50\times10^{-5}$ & $7.50\times10^{-5}$   \\
O       & $1.76\times10^{-4}$ & $1.76\times10^{-4}$  & $3.19\times10^{-4}$ & $3.19\times10^{-4}$   \\
F       & $6.68\times10^{-9}$ & $6.68\times10^{-9}$  & --		   & -- 		   \\
C$^+$   & $7.30\times10^{-5}$ & $7.30\times10^{-5}$  & $1.32\times10^{-4}$ & $1.32\times10^{-4}$   \\
Na$^+$  & $2.00\times10^{-9}$ & $2.00\times10^{-9}$  & --		   & -- 		   \\
Si$^+$  & $8.00\times10^{-9}$ & $8.00\times10^{-9}$  & --		   & -- 		   \\
S$^+$   & $8.00\times10^{-8}$ & $8.00\times10^{-8}$  & $1.86\times10^{-5}$ & $1.86\times10^{-5}$   \\
Mg$^+$  & $7.00\times10^{-9}$ & $7.00\times10^{-9}$  & --		   & -- 		   \\
P$^+$   & $3.00\times10^{-9}$ & $3.00\times10^{-9}$  & --		   & -- 		   \\
Cl$^+$  & $4.00\times10^{-9}$ & $4.00\times10^{-9}$  & --		   & -- 		   \\
Fe$^+$  & $3.00\times10^{-9}$ & $3.00\times10^{-9}$  & $1.50\times10^{-8}$ & $1.50\times10^{-8}$   \\
H$_2$O  & --                  & $5.00\times10^{-5}$  & --		   & -- 		   \\
H$_2$CO & --                  & $2.00\times10^{-6}$  & --		   & -- 		   \\
CH$_4$O & --                  & $2.00\times10^{-6}$  & --		   & -- 		   \\
CH$_4$  & --                  & $5.00\times10^{-7}$  & --		   & -- 		   \\
			       
Temperature (K) & 10 & 100 & 20 & 90\\
H$_2$ density (cm$^{-3})$ & $1\times10^{4}$ & $1\times10^{7}$ & $25$ & $5\times10^{4}$ \\
Cosmic ray ionization rate & $1.3\times10^{-17}$ &  $1.3\times10^{-17}$ &  $5.0\times10^{-17}$&  $5.0\times10^{-17}$ \\
Visual extinction & 10 & 10 & 1 & 10
\enddata
\tablenotetext{1}{The elemental abundances are those at the beginning of the
simulation, while the molecular abundances (H$_2$O, H$_2$CO, CH$_4$O, and
CH$_4$) are set to the given values at a model time of $10^5$~years.}
\tablecomments{There is no F included in the OSU gas-grain code.}
\end{deluxetable}

\begin{deluxetable}{ccccccc}
\tablewidth{0pt}
\tablecaption{Maximum abundance differences found when including DR in
the OSU gas-grain code relative to excluding DR.}
\tablehead{\colhead{} & \colhead{} & \colhead{Maximum} & 
\colhead{Time} & \colhead{Relative}& 
\colhead{Gas-Phase} & \colhead{Surface} \\
\colhead{Species} & \colhead{Phase} & \colhead{Difference} & 
\colhead{(years)} & \colhead{Abundance}& 
\colhead{Detection\tablenotemark{1}} & \colhead{Detection\tablenotemark{2}}}
\startdata
\label{tab:abundances}

        O$_3$ & Surface &1440  &      		$1.78\times10^4$  &  $1.06\times10^{-8}$  &   No  &   No  \\
      N$_2$H$_2$ & Surface &  59.3  &      	$3.16\times10^4$  &  $1.47\times10^{-12}$ &   No  &   No  \\
      H$_2$O$_2$ & Surface &  45.0  &     	$5.62\times10^4$  &  $7.23\times10^{-9}$  &   No  &   No  \\
      HNCO &     Gas &  38.7  &      		$1.78\times10^4$  &  $6.13\times10^{-12}$ &   Yes &   No  \\
      HNCO & Surface &  37.0  &      		$1.78\times10^4$  &  $2.93\times10^{-9}$  &   Yes &   No  \\
        NO & Surface &  34.9  &      		$3.16\times10^4$  &  $5.44\times10^{-12}$ &   Yes &   No  \\
    NH$_2$CHO & Surface &  30.0  &      	$5.62\times10^4$  &  $3.46\times10^{-11}$ &   Yes &   No  \\
       C$_2$H & Surface &  27.2  &      	$1.00\times10^4$  &  $1.06\times10^{-12}$ &   Yes &   No  \\
      C$_2$H$_4$ & Surface &  23.1  &      	$1.00\times10^4$  &  $5.64\times10^{-10}$ &   Yes &   No  \\
      HC$_2$O &     Gas &  18.3  &      	$1.78\times10^4$  &  $2.41\times10^{-12}$ &   No  &   No  \\
     CH$_2$CO & Surface &  15.0  &      	$1.00\times10^4$  &  $3.10\times10^{-10}$ &   No  &   No  \\
     CH$_2$OH &     Gas &  14.1  &      	$1.00\times10^4$  &  $5.47\times10^{-12}$ &   No  &   No  \\
     CH$_3$OH &     Gas &  13.8  &      	$1.00\times10^4$  &  $3.26\times10^{-12}$ &   Yes &   Yes \\
     C$_3$H$_3$N & Surface &  12.8  &   	$3.16\times10^4$  &  $1.98\times10^{-12}$ &   No  &   No  \\
      HC$_3$N & Surface &  12.5  &      	$3.16\times10^4$  &  $3.77\times10^{-11}$ &   Yes &   No  \\
      C$_2$H$_6$ &     Gas &  11.3  &      	$1.00\times10^4$  &  $5.07\times10^{-12}$ &   No  &   No  \\
    CH$_2$NH$_2$ &     Gas &   9.76  &      	$5.62\times10^4$  &  $1.96\times10^{-12}$ &   No  &   No  \\
     CH$_3$NH &     Gas &   9.76  &      	$5.62\times10^4$  &  $3.65\times10^{-12}$ &   No  &   No  \\
      CH$_5$N & Surface &   9.55  &      	$5.62\times10^4$  &  $9.48\times10^{-9}$  &   No  &   No  \\
      C$_2$H$_2$ & Surface &   8.27  &      	$1.00\times10^4$  &  $1.19\times10^{-9}$  &   Yes &   No  \\
     CH$_3$OH & Surface &   7.59  &      	$3.16\times10^4$  &  $2.43\times10^{-8}$  &   Yes &   No  \\
      N$_2$H$_2$ &     Gas &   7.44  &      	$1.00\times10^5$  &  $5.57\times10^{-12}$ &   No  &   No  \\
     CH$_3$CN & Surface &   7.39  &      	$1.00\times10^4$  &  $7.14\times10^{-11}$ &   Yes &   No  \\
       HCN & Surface &   7.37  &      		$1.78\times10^4$  &  $1.24\times10^{-8}$  &   Yes &   No  \\
       CH$_3$ & Surface &   6.96  &      	$5.62\times10^4$  &  $4.42\times10^{-12}$ &   Yes &   No  \\
       CH$_2$ & Surface &   6.96  &      	$5.62\times10^4$  &  $4.46\times10^{-12}$ &   Yes &   No  \\
       HNC & Surface &   6.82  &      		$1.78\times10^4$  &  $6.08\times10^{-9}$  &   Yes &   No  \\
        CH & Surface &   6.74  &      		$5.62\times10^4$  &  $3.99\times10^{-12}$ &   Yes &   Maybe  \\
         C & Surface &   6.74  &      		$5.62\times10^4$  &  $4.03\times10^{-12}$ &   No  &   No  \\
      C$_2$H$_6$ & Surface &   6.63  &      	$3.16\times10^4$  &  $1.25\times10^{-8}$  &   No  &   No  \\
         N & Surface &   6.51  &      		$5.62\times10^4$  &  $1.99\times10^{-12}$ &   No  &   No  \\
        NH & Surface &   6.51  &      		$5.62\times10^4$  &  $2.02\times10^{-12}$ &   Yes &   No  \\
       NH$_2$ & Surface &   6.50  &      	$5.62\times10^4$  &  $2.09\times10^{-12}$ &   Yes &   No  \\
         O & Surface &   6.40  &      		$5.62\times10^4$  &  $1.48\times10^{-11}$ &   No  &   No  \\
      C$_4$H$_2$ & Surface &   6.36  &      	$5.62\times10^4$  &  $6.67\times10^{-11}$ &   Yes &   No  \\
      C$_9$H$_2$ & Surface &   6.35  &      	$5.62\times10^4$  &  $1.00\times10^{-12}$ &   No  &   No  \\
      H$_2$CO & Surface &   6.05  &      	$1.78\times10^4$  &  $3.75\times10^{-8}$  &   Yes &   Maybe  \\
     H$_2$C$_3$O &     Gas &   5.88  &      	$3.16\times10^4$  &  $1.01\times10^{-12}$ &   Yes &   No  \\
      HC$_3$O &     Gas &   5.88  &      	$3.16\times10^4$  &  $1.90\times10^{-12}$ &   No  &   No  \\
     H$_2$C$_3$O & Surface &   5.86  &      	$3.16\times10^4$  &  $3.53\times10^{-10}$ &   Yes &   No  \\
      C$_3$H$_2$ & Surface &   5.70  &      	$5.62\times10^4$  &  $1.50\times10^{-8}$  &   Yes &   No  \\
        OH & Surface &   5.49  &      		$5.62\times10^4$  &  $1.24\times10^{-11}$ &   Yes &   No  \\
      C$_5$H$_2$ & Surface &   5.43  &      	$5.62\times10^4$  &  $6.92\times10^{-12}$ &   No  &   No  \\
     CH$_2$NH &     Gas &   4.89  &      	$1.00\times10^4$  &  $4.16\times10^{-12}$ &   Yes &   No  \\
        O$_2$ & Surface &   4.62  &      	$1.78\times10^4$  &  $3.97\times10^{-10}$ &   Yes &   No  \\
        O$_3$ &     Gas &   4.25  &      	$1.00\times10^5$  &  $6.71\times10^{-10}$ &   No  &   No  \\
     H$_5$C$_3$N & Surface &   3.75  &      	$3.16\times10^4$  &  $1.03\times10^{-10}$ &   No  &   No  \\
       N$_2$O &     Gas &   3.48  &      	$1.00\times10^4$  &  $5.05\times10^{-11}$ &   Yes &   No  \\
       H$_2$S & Surface &   3.41  &      	$5.62\times10^4$  &  $5.08\times10^{-9}$  &   Yes &   No  \\
      H$_2$O$_2$ &     Gas &   3.37  &      	$3.16\times10^4$  &  $9.22\times10^{-11}$ &   No  &   No  \\
      CH$_5$N &     Gas &   3.26  &      	$1.00\times10^4$  &  $2.19\times10^{-12}$ &   No  &   No  \\
       N$_2$O & Surface &   3.19  &      	$1.78\times10^4$  &  $1.40\times10^{-12}$ &   Yes &   No  \\
       NO$_2$ &     Gas &   3.04  &      	$1.00\times10^4$  &  $2.27\times10^{-11}$ &   No  &   No  \\
       HNO &     Gas &   2.98  &      		$1.00\times10^4$  &  $2.97\times10^{-10}$ &   Yes &   No  \\
       HNO & Surface &   2.76  &      		$1.00\times10^4$  &  $8.76\times10^{-9}$  &   Yes &   No  \\
     CH$_3$CN &     Gas &   2.71  &      	$1.00\times10^4$  &  $6.84\times10^{-12}$ &   Yes &   No  \\
      C$_6$H$_2$ & Surface &   2.69  &      	$3.16\times10^4$  &  $1.17\times10^{-12}$ &   Yes &   No  \\
       NO$_2$ & Surface &   2.62  &      	$3.16\times10^4$  &  $1.05\times10^{-12}$ &   No  &   No  \\
       C$_3$H & Surface &   2.44  &      	$1.00\times10^4$  &  $1.02\times10^{-12}$ &   Yes &   No  \\
       CO$_2$ & Surface &   2.23  &      	$1.00\times10^4$  &  $2.08\times10^{-10}$ &   Yes &   Yes \\
    CH$_3$CO$^+$ &     Gas &   2.16  &      	$1.00\times10^4$  &  $2.59\times10^{-12}$ &   No  &   No  \\
     HC$_2$NC &     Gas &   2.09  &      	$1.00\times10^4$  &  $1.51\times10^{-12}$ &   Yes &   No  \\
     C$_3$H$_3$$^+$ &     Gas &   2.09  &      	$1.00\times10^4$  &  $2.55\times10^{-12}$ &   No  &   No  \\
      HC$_3$N &     Gas &   2.02  &      	$1.00\times10^4$  &  $2.99\times10^{-11}$ &   Yes &   No  \\
&&&&&\\
      H$_2$CS & Surface &   0.141  &      	$1.00\times10^4$  &  $1.93\times10^{-12}$ &   Yes &   No  \\
       C$_2$S & Surface &   0.210  &      	$5.62\times10^4$  &  $1.25\times10^{-12}$ &   Yes &   No  \\
      C$_5$H$_4$ & Surface &   0.292  &      	$1.00\times10^4$  &  $2.82\times10^{-12}$ &   No  &   No  \\
      C$_3$H$_4$ & Surface &   0.323  &      	$1.00\times10^4$  &  $4.01\times10^{-9}$  &   No  &   No  \\
        CS & Surface &   0.325  &      		$1.00\times10^4$  &  $1.71\times10^{-12}$ &   Yes &   No  \\
      C$_6$H$_4$ & Surface &   0.349  &      	$1.78\times10^4$  &  $4.65\times10^{-13}$ &   No  &   No  \\
       HS$_2$ &     Gas &   0.384  &      	$5.62\times10^4$  &  $6.88\times10^{-12}$ &   No  &   No  \\
      H$_2$S$_2$ &     Gas &   0.386  &      	$5.62\times10^4$  &  $6.99\times10^{-12}$ &   No  &   No  \\
       H$_2$S &     Gas &   0.392  &      	$3.16\times10^4$  &  $1.80\times10^{-10}$ &   Yes &   No  \\
      H$_3$S$^+$ &     Gas &   0.413  &      	$5.62\times10^4$  &  $7.64\times10^{-13}$ &   No  &   No

\enddata
\tablecomments{We list the maximum difference factor 
in abundance when DR is included in the OSU gas-grain code relative to when 
it is left out.  At the relevant time in the model evolution, we list the 
abundance of each species
relative to total H nuclei density when DR is included.  We also list whether there has been an interstellar
detection of the molecule in the gas phase or on dust surfaces.}
\tablenotetext{1}{http://astrochymist.org/astrochymist\_ism.html}
\tablenotetext{2}{\citet{Gibb04a}}
\end{deluxetable}

\end{document}